\documentclass[prl,aps,showpacs,preprintnumbers,twocolumn]{revtex4}
\usepackage{graphics,bm}

\def\bx{{\bf x}}
\def\bk{{\bf k}}

\begin{document}

\title{The bosonic Kondo effect}

\author{G.M. Falco}
\author{R.A. Duine}
\author{H.T.C. Stoof}
\affiliation{Institute for Theoretical Physics,
         University of Utrecht, Leuvenlaan 4,
         3584 CE Utrecht, The Netherlands}
\date{\today}

\begin{abstract}
The Kondo effect is associated with the formation of a many-body
ground state that contains a quantum-mechanical entanglement
between a (localized) fermion and the free fermions. We show that
a bosonic version of the Kondo effect can occur in degenerate atomic
Fermi gases near a Feshbach resonance. We also discuss how this bosonic Kondo effect
can be observed experimentally.
\end{abstract}

\pacs{03.75.Kk, 67.40.-w, 32.80.Pj}
\preprint{ITP-UU-03/22;cond-mat/0304489}
\maketitle
\def\bx{{\bf x}}
\def\bk{{\bf k}}
\def\half{\frac{1}{2}}
\def\args{(\bx,t)}

{\it Introduction.} --- The Kondo effect is an intricate many-body
phenomenon, that was originally put forward to explain the
anomalous resistance minimum of metals contaminated with magnetic
impurities. Today the Kondo effect is known to be also responsible
for the existence of heavy-fermion materials \cite{Hewson}, and is
speculated to play an important role in the physics of
high-temperature superconductors \cite{Cox}. Moreover, a
particularly clear manifestation of this effect occurs in
semiconductor quantum dots \cite{Gordon,Cronen,Sasa}. In all these
examples the Kondo effect is associated with the formation of a
many-body ground state that contains a quantum-mechanical
entanglement between a (localized) fermion and the free fermions.
Here we show that a new version of the Kondo effect
can occur in the degenerate atomic Fermi gases that have recently
been created \cite{Demarco,Trusco,Schre,Modu,Dieck}. In contrast
to the fermionic version, the many-body ground state of the gas
shows now coherence between bosons, in fact bosonic molecules, and
the free fermions in the gas. We explore the conditions under which
this bosonic Kondo effect occurs and also discuss how it can be
observed experimentally.

To be concrete we consider a degenerate Fermi gas of $^{6}$Li or
$^{40}$K atoms near a so-called Feshbach resonance
\cite{Dieck,O'Ha,Regal,Bourdel}. A Feshbach resonance in the
scattering amplitude of two alkali atoms arises when the energy of
the two colliding atoms is close to the energy of a diatomic
molecular state that is coupled to the incoming atoms. Since the
coupling is provided by the hyperfine interaction between the
electron and nuclear spins of the atoms, the magnetic moment of
the molecule is not equal to twice the atomic magnetic moment. As
a result, the Zeeman interaction leads to an energy difference between
the incoming atoms and the molecule, and therefore to an effective
interaction strength between the atoms, that is tunable by
means of an external magnetic field. The usefulness of these
Feshbach resonances for degenerate atomic gases was first pointed
out theoretically \cite{Stwa,Tie} and then confirmed
experimentally in a Bose-Einstein condensate of $^{23}$Na atoms
\cite{Ino}.  
Feshbach resonances 
appear to be particularly useful for
ultracold Fermi gases because of the possibility to achieve a
Bose-Einstein condensation of Cooper pairs at experimentally
accessible temperatures in this way \cite{Stoof,Com,Pet,Oh,Hol,Mil}. For our purposes
it is most important to realize that, due to the nearby location
of a molecular state, the atomic gas can develop coherence between
atoms and molecules. In a degenerate Bose gas the atom-molecule
coherence has last year been demonstrated experimentally by Donley
{\it et al}. \cite{Cla}. It is the purpose of this Letter to point out
that in the case of a degenerate Fermi gas it leads to Kondo
physics.

{\it Atom-molecule model.} --- In a dilute gas 
at low temperature
only $s$-wave scattering is important.  
In a gas of fermionic particles, 
resonant interaction effects 
can therefore only be observed if we 
consider a mixture of two hyperfine states, denoted by
$|\!\uparrow\rangle$ and $|\!\downarrow\rangle$. Close to
resonance the gas is then described by an effective
atom-molecule action \cite{Du}.
In this action atoms in different hyperfine states 
are coupled to dressed
molecular bosons with twice the atomic mass $m$. 
In the absence of
the coupling, the energy of the molecules relative to the
two-atom continuum threshold is equal to the magnetic-field
dependent detuning energy $\delta(B)$. The coupling terms describe
processes in which two fermionic atoms form one bosonic molecule
or the bosonic molecule breaks up into two fermionic atoms. The
effective coupling strength $g$ corresponds physically to the bare
atom-molecule coupling dressed with ladder diagrams to take into
account all two-body processes. According to this renormalization procedure, 
the
parameters $g$ and $\delta(B)$ of the effective atom-molecule
theory can be expressed solely in terms of the experimentally
known location of the Feshbach resonance 
$B_0$, its width $\Delta B$ and the
associated magnetic moment difference $\Delta \mu $. 
In detail we have $g=\hbar\sqrt{4\pi\Delta B\Delta\mu/m}$ and
$\delta(B)=\Delta\mu(B-B_0)$ \cite{Du}.
\begin{figure}
\includegraphics{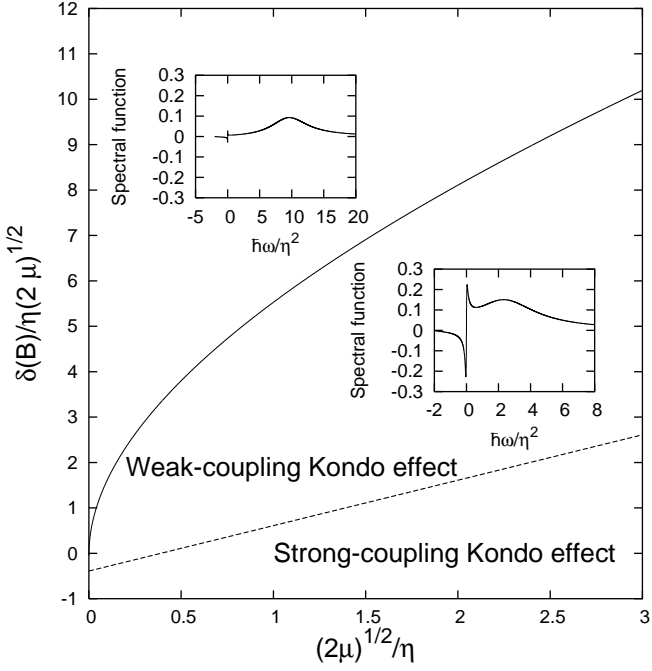}
\caption{The phase diagram for the bosonic Kondo effect. Above the
solid line the Kondo effect is for all practical purposes
invisible, whereas below this line there appears a Kondo resonance
in the molecular spectral function. The dashed line marks the
crossover to the strong-coupling regime. In the insets we show the
zero-momentum molecular spectral function in units of $1/\eta^2$
and as a function of frequency, for two different values of the
parameters $\delta(B)/\eta\sqrt{2\mu}$ and $\sqrt{2\mu}/\eta$. }
\label{fig1}
\end{figure}
We consider here only the case of an equal number of atoms in
every hyperfine state and hence there is only one chemical $\mu$.
To avoid Bose-Einstein
condensation of the molecules we consider a positive detuning that
is larger than twice the chemical potential $\mu$ of the atoms. In
this case the molecule has a finite lifetime because it can decay
into two free atoms. This finite lifetime is incorporated by the
imaginary part of the molecular self-energy.

{\it Spectral weight function.} --- Because we are mostly
interested in the density of states of the molecules, we focus
here on the spectral properties of the molecular Green's function
and in particular how they are affected by the presence of the
Fermi sea of atoms. According to quantum many-body theory the
molecular density of states is related to the spectral function
$\rho_{\rm m}(\bf q,\omega)$, which is found from the retarded
Green's function or propagator by means of $ \rho_{\rm m}({\bf
q},\omega)=-{\rm Im}[G_{\rm m} ^ {(+)}({\bf
q},\omega)]/\pi\hbar$ and obeys the sum rule $ \int_{-\infty} ^{\infty}
d(\hbar \omega)\,\,\rho_{\rm m}({\bf q},\omega)=1$. The exact
molecular retarded propagator is
\begin{equation}
  G_{{\rm m}}^{(+)}({\bf q},\omega)=\frac{\hbar}{\hbar
  \omega+2\mu-\delta(B) -\epsilon_{{\bf q}}/2-\hbar \Sigma_{\rm m}^
  {(+)}({\bf q},\omega)},
\end{equation} 
where $\epsilon_{\bf q}=\hbar^2{\bf q}^2/2m$
and $\hbar \Sigma ^ {(+)} _{\rm m}({\bf q},\omega)$
is the exact retarded selfenergy. The problem thus reduces to the
evaluation of the selfenergy of the molecule. In the presence of a
degenerate Fermi gas at temperatures far below the Fermi
temperature, a zero-temperature approximation to the Fermi gas can
be adopted. For the low-energy Kondo physics of interest to us the
energy dependence of the coupling constant $g$ can be negleted as
well. Moreover, we have checked that, within the weak-coupling
limit of the Kondo effect that we only consider here, it is
appropriate to consider the fermionic propagators as free and
neglect mean-field effects due to interactions between the
fermions in different spin states. In the ladder approximation the
selfenergy is determined by a Feynman diagram that physically
describes the process of a dressed molecule decaying into two
atoms and subsequently recombining into a molecule again. Summing
over all ladder diagrams ultimately yields
\begin{widetext}
\begin{eqnarray}
\hbar\Sigma_{{\rm m}}^{(+)}({\bf q},\omega)&&=-\eta i
\sqrt{\hbar\omega+2\mu-\frac{\epsilon_{{\bf
q}}}{2}}+2\eta\frac{\sqrt{2 \mu}}{\pi}+\eta \frac{\hbar\omega}{\pi\sqrt{2\epsilon_{\bf q}}}
{\rm{ln}}\left[\frac{\hbar\omega-\epsilon_{\bf q}+2\sqrt{\mu\epsilon_{\bf q} }}
{\hbar\omega-\epsilon_{\bf q}-2\sqrt{\mu\epsilon_{\bf q} }}
\right]+\nonumber\\
&&+\frac{\eta}{\pi}\sqrt{\hbar\omega+2\mu-\frac{\epsilon_{{\bf q}}}{2}}\left[
{\rm{ln}}\frac{\sqrt{\hbar\omega+2\mu-\frac{\epsilon_{{\bf q}}}{2}}
-(\sqrt{2\mu}+\sqrt{\frac{\epsilon_{{\bf q}}}{2}})}
{\sqrt{\hbar\omega+2\mu-\frac{\epsilon_{{\bf q}}}{2}}
+(\sqrt{2\mu}+\sqrt{\frac{\epsilon_{{\bf q}}}{2}})}+
{\rm{ln}}\frac{\sqrt{\hbar\omega+2\mu-\frac{\epsilon_{\bf q}}{2}}
-(\sqrt{2\mu}-\sqrt{\frac{\epsilon_{{\bf q}}}{2}})}
{\sqrt{\hbar\omega+2\mu-\frac{\epsilon_{{\bf q}}}{2}}
+(\sqrt{2\mu}-\sqrt{\frac{\epsilon_{{\bf q}}}{2}})}
\right],
\end{eqnarray}
\end{widetext}
where $\eta=g^2m^{\frac{3}{2}}/4\pi\hbar^3$.

The bosonic spectral density for two different values of the
detuning $\delta(B)$ is shown in the insets of Fig.~\ref{fig1}. In
the upper inset $\delta(B)/\eta\sqrt{2\mu}=12/\sqrt{2}$ and
$\sqrt{2\mu}/\eta=\sqrt2$ so that the detuning is much larger than
the Fermi energy, or more precisely $\delta\gg\eta\sqrt{2\mu}$. In
that case there is essentially no Kondo effect and the spectral
density shows just a single broad peak centered around the
detuning. This is the expected situation for a single molecular
state with a finite lifetime. In the lower inset, where
$\delta(B)/\eta\sqrt{2\mu}=5/\sqrt2$ and $\sqrt{2\mu}/\eta=\sqrt2$
the detuning is much closer to twice the Fermi energy and, apart
from the broad feature around the detuning, the spectral density
now also shows slightly above and below zero frequency two sharp
peaks. This is the Kondo effect. Physically these two sharp peaks are a
pure many-body effect due to a logarithmic
singularity in the selfenergy that is induced by the Fermi surface. They signal
the formation of a molecular resonance that contains a
quantum-mechanical entanglement between the bosonic molecule and
the free fermions. The situation is reminiscent of the Anderson model for
a quantum dot, with a localized electron
level located just below the Fermi energy of the leads. In that
model the onsite Coulomb repulsion on the quantum dot
removes the symmetry between scattering to and from the dot and
leads to a logarithmic singularity in the selfenergy. In our case
no Coulomb blockade is required and the formation of the Kondo
resonance is the result of the fact that the molecule consists of
two atoms. If a molecule breaks apart it adds two particles to the
Fermi sea. As a result our Hamiltonian is not quadratic in
creation and annihilation operators and contains much more physics
than the Anderson model without onsite Coulomb interaction, {\it i.e.}, the so
called Fano-Anderson model which has no Kondo resonance.
The role of the localized fermion is
now played by the bosonic molecular state above the two-particle
Fermi surface and the Kondo resonance appears slightly above twice
the Fermi energy because otherwise the equilibrium situation would be a
Bose-Einstein condensate of molecules.

In the weak-coupling limit the molecular state with zero momentum 
has exactly the same form as the wave-function underlying the
Kondo effect in the Anderson model.
In the latter case we have \cite{Mahan}
\begin{eqnarray}
|\Psi_{\rm K,\sigma}\rangle=\sqrt{Z} f_{\sigma}^{\dagger}|\Phi_{\rm
F}\rangle+\sum_{ |\bk|> k_{\rm F}} u_{\bf k}c_{\bf
k,\sigma}^{\dagger}|\Phi_F\rangle+\nonumber\\
\sum_{ |\bk|< k_{\rm F}}v_{\bf k}f_{\sigma}^{\dagger}f_{-\sigma
}^{\dagger} c_{\bf k,-\sigma}|\Phi_{\rm F}
\rangle\,\,\,,
\end{eqnarray} where $|\Phi_{\rm F}\rangle$ is the
filled Fermi sphere, $k_{\rm F}$ is the Fermi momentum and, 
$c_{\bf k,\sigma}^{\dagger}$ and $ f_{\sigma}^{\dagger}$ are the
creation operators for the conduction electrons
and the electrons on the dot, respectively.
The only difference with the Anderson model is that
in our situation fermions are created above and below the Fermi sea in pairs.
Hence we have
\begin{eqnarray}
|\Psi_{\rm K}\rangle=\sqrt{Z} b_{\bf 0}^{\dagger}|\Phi_{\rm
F}\rangle+\sum_{|\bk|> k_{\rm F}} u_{\bf k}c_{\bf
k,\uparrow}^{\dagger}c_{\bf
-k,\downarrow}^{\dagger}|\Phi_F\rangle+\nonumber\\
\sum_{|\bk|< k_{\rm F}}v_{\bf k}b_{\bf 0}^{\dagger}b_{\bf
0}^{\dagger} c_{\bf k,\uparrow}c_{-\bf k,\downarrow}|\Phi_{\rm F}
\rangle\,\,\,,
\end{eqnarray} where 
$b_{\bf 0}^{\dagger}$ and $c_{\bf k,\sigma}^{\dagger}$ are
our bosonic and fermionic
creation operators.
In the
two-atom limit, where the Fermi sea $|\Phi_{\rm F}\rangle$ becomes
the vacuum state and no Kondo effect is present, the hole
amplitude $v_{\bf k}$ vanishes and the molecular wave function
only has contributions from the bare molecular state, with
amplitude $\sqrt{Z}$, and from the two-atom continuum, with
particle amplitudes $u_\bk$. This wave function indeed describes
exactly the dressed molecule of the Feshbach resonance. When the
Fermi surface is present, pairs of fermionic atoms with opposite
spin tunnel back and forth from  the Fermi sea into the molecular state. The
coherent addition of many of such events produces the Kondo
effect. It is important to realize that because the molecules are
bosons, the negative peak below the Fermi energy and the positive
peak above the Fermi energy are physically both associated with
the same molecular state. The presence of the negative peak just
indicates that molecular states are populated even at zero
temperature.

{\it Kondo temperature.} --- The energy of the Kondo resonance is
equal to $k_{\rm B} T_{\rm K}$, where $T_{\rm K}$ is the so-called
Kondo temperature. This temperature plays an important role
because in the Kondo regime it is possible to express all the
temperature effects as universal functions of $T/T_{\rm K}$.
Moreover, it is the temperature to which the Fermi gas needs to be
cooled experimentally to observe the Kondo effect. In the
weak-coupling limit it is given by
\begin{equation}
\label{eq:tkondo}
 T_{\rm K}=\frac{8\mu}{e^2k_{\rm B}}e^{\frac{\pi}{2}
\frac{\sqrt{2\mu}}{\eta}}e^{-\frac{\pi}{2}\frac{\delta(B)}{\eta\sqrt{2\mu}}}\,\,.
\end{equation}

\noindent This non-perturbative result is, 
apart from the prefactor, identical to the
Kondo temperature of the Anderson model in the weak coupling limit
that is given by $
 T_{\rm K} \propto (\mu/k_{\rm B})
 e^{-\delta/\rm g^2 N(0)}$,
where $\delta$ is the energy difference between the state on
the quantum dot and the Fermi level, $\rm g$ is the hopping parameter
between the conduction electrons and the dot, and $N(0)$
is the density of states at the Fermi level.

The Kondo temperature makes it possible to classify two different
limits. When $k_{\rm B} T_{\rm K} \ll 2\mu$ the system is in the
weak-coupling regime. Otherwise we are dealing with a
strong-coupling situation. The dashed line in the phase diagram
shown in Fig.~\ref{fig1} marks this separation in the parameter
space. The theory presented here allows for a full description of
the weak-coupling limit, while for a strong-coupling treatment an
improved approximation in the calculation of the molecular
selfenergy is needed. In particular, it will be necessary to
consider the resonant interactions between the fermions, which
lead to the creation of particle-hole pairs in the Fermi sea.

{\it Experimental signature.} --- In order to observe the Kondo
effect, its associated resonance must contain sufficient spectral
weight. Using that the integrated weight of  the Kondo peaks in
the spectral function is proportional to $k_{\rm B}
T_{\rm K}/\eta\sqrt{2\mu}$, the solid line in Fig.~\ref{fig1} gives a
quantitative estimate of the lower limit for the visibility of the
effect. Using typical parameters of $^6$Li and $^{40}$K, we have
checked that the Kondo regime is indeed experimentally accessible.
The direct experimental measurement of this Kondo effect can be
accomplished by looking at the enhancement in the density of
molecules even for relatively large detuning. 
For example, the recently developed method for
sensitive detection of molecules in cold atomic gases
could be used for this purpose \cite{Chi}.
 
\begin{figure}[h]
\includegraphics{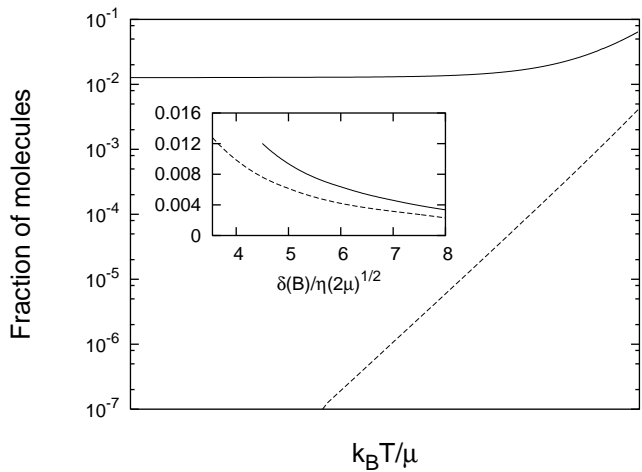}
\caption{Fraction of molecules in the gas as a function of
temperature. The solid line shows the result for the parameter
$\delta(B)/\eta\sqrt{2\mu}=5/\sqrt2$ and
$\sqrt{2\mu}/\eta=\sqrt2$. The dashed line shows 
the maximum number of molecules that can be achived without many-body effects.
In the inset the
fraction of molecules
at zero temperature is plotted as a function of the detuning for
two different values of the chemical potential. The solid 
and the dashed lines show
the result when $\sqrt{2\mu}/\eta=2$  
and when $\sqrt{2\mu}/\eta=\sqrt2$, respectively. 
} \label{fig2}
\end{figure}

The influence of the Kondo effect on the total number of molecules
in the gas can be easily calculated. At temperatures $T\ll\mu/k_{\rm B}$
it is consistent to calculate the total density of the bosonic
molecules in the gas by multiplying the zero-temperature molecular
spectral function with the Bose distribution function
$N(x)=[e^x-1]^{-1}$ and integrating over all momenta and
frequencies, i.e., 
\begin{equation}
n_{\rm B}(T)=\int_{-\infty}^{+\infty}\,d(\hbar\omega)\int \frac{{\rm d}
{\bf q}}{(2\pi)^3}\, \rho_{\rm m}({\bf q}
,\omega)N\left(\frac{\hbar\omega}{k_{\rm B} T}\right).
\end{equation} 
At zero
temperature and without many-body effects, the number of molecules
is exactly zero because the detuning is larger than the Fermi
energy. When the Kondo resonance is included, even at zero
temperature a nonzero number of molecules is obtained. The inset of 
Fig.~\ref{fig2} shows the molecular fraction in the gas at zero
temperature for different values of the detuning.
The thermal effects on the number of
molecules are also shown in Fig.~\ref{fig2}. Notice the enormous
enhancement in the number of molecules due to the Kondo effect. It
is also interesting to observe that without the many-body effects
the maximum number of molecules goes to zero as $T^3$, whereas with the
Kondo effect it saturates to a nonzero constant only as $T^2$.
Since
our calculations are valid within the weak-coupling limit, results
are shown only for this regime.

It is important to realize that the Kondo temperature found above is
always larger than the critical temperature of the Bose-Einstein
condensation of Cooper pairs and, therefore, more easy to obtain
experimentally. In the weak-coupling limit this critical
temperature is known in terms of the s-wave scattering length $a({\rm B})$
\cite{Stoof} as $
 T_{\rm{BCS}}=\left(8 e^{\gamma-2}\mu/\pi k_{\rm B}\right)
e^{\frac{\pi}{2k_F a(B)}}$, where $\gamma$ is Euler's constant.
 Using the fact that the resonant part of this scattering length is given by
$-g^2m/4\pi\hbar^2\delta(B)$, allows us to express the Kondo temperature in terms of the BCS critical
temperature as $
T_{\rm K}=\pi
e^{\frac{\pi}{2}\frac{\sqrt{2\mu}}{\eta}-\gamma}\,T_{\rm {BCS}}$,
which is indeed always larger than the BCS critical temperature.
Nevertheless, it must pointed out that since only the parameters
of the resonant part of the interaction play a role for the Kondo effect,
$T_{\rm K}$ is physically quite distinct from $T_{\rm BCS}$. In particular, we can have the
situation of a positive total scattering length with no Cooper pairing at
any temperature, but with the Kondo effect still present.

\begin{acknowledgments}
We thank  M. Baranov, M. Cazalilla, G. 't Hooft, R. Hulet, B. Farid and S. Stringari for helpful remarks.
This work is supported by the Stichting voor Fundamenteel Onderzoek der Materie
(FOM) and the Nederlandse
Organisatie voor Wetenschappelijk Onderzoek (NWO).
\end{acknowledgments}


\begin{thebibliography}{99}
\bibitem{Hewson}
A. C. Hewson, {\it {The Kondo Problem to Heavy Fermions}} (Cambridge University Press, Cambridge, 1993).
\bibitem{Cox}
D. L. Cox, and M. B. Maple, Phys. Today \textbf{48}, 32 (1995).
\bibitem{Gordon}
D. Goldhaber-Gordon {\it et al.},  Nature \textbf{391}, 156 (1998).
\bibitem{Cronen}
S. M. Cronenwett, T. H. Oosterkamp, and L. P. Kouwenhoven, Science \textbf{281}, 540 (1998).
\bibitem{Sasa}
S. Sasaki {\it et al.}, Nature \textbf{405}, 764 (2000).
\bibitem{Demarco}
B. DeMarco, and D. Jin, Science \textbf{285}, 1703 (1999).
\bibitem{Trusco}
A. G. Truscott, K. E. Strecker, W. I. McAlexander, G. B. Partridge, and R. G. Hulet,
Science \textbf{291}, 2570 (2001).
\bibitem{Schre}
F. Schreck {\it et al.}, Phys. Rev. Lett. \textbf{87}, 080403 (2001).
\bibitem{Modu}
G. Modugno {\it et al.}, Science \textbf{297}, 2240 (2002).
\bibitem{Dieck}
K. Dieckmann {\it et al.},
Rev. Lett. \textbf{89}, 203201 (2002).
\bibitem{O'Ha}
K. M. O'Hara, S. L.  Hemmer, M. E. Gehm, S. R. Granade, and J. E. Thomas, 
Science \textbf{298}, 2179 (2002).
\bibitem{Regal}
C. A. Regal and D. S. Jin,
Preprint cond-mat/0302246.
\bibitem{Bourdel}
T. Bourdel {\it et al.},
Preprint cond-mat/0303079.
\bibitem{Stwa}
W. C. Stwalley, Phys. Rev. Lett. \textbf{37}, 1628 (1976).
\bibitem{Tie}
E. Tiesinga, B. J. Verhaar, and H. T. C. Stoof,
Phys. Rev. A \textbf{47}, 4114 (1993).
\bibitem{Ino}
S. Inouye {\it et al.}, Nature \textbf{392}, 151 (1998).
\bibitem{Stoof}
H.T.C. Stoof,  M. Houbiers,  C.A. Sackett, and R.G. Hulet,
Phys. Rev. Lett. {\bf 76}, 10 (1996).
\bibitem{Com}
R. Combescot, Phys. Rev. Lett. {\bf 83}, 3766 (1999).
\bibitem{Pet}
H. Heiselberg, C. J. Pethick, H. Smith, and L. Viverit, Phys. Rev. Lett. {\bf 85}, 2418 (2000).
\bibitem{Hol}
M. Holland, S. J. J. M. F. Kokkelmans,  M. L. Chiofalo, and R. Walser, 
Phys. Rev. Lett. \textbf{87}, 120406 (2001).
\bibitem{Mil}
J. N. Milstein, S. J. J. M. F. Kokkelmans, and M. J. Holland, 
Phys. Rev. A {\bf 66}, 043604 (2002).
\bibitem{Oh}
Y. Ohashi, and  A. Griffin, Phys. Rev. Lett.
\textbf{89}, 130402 (2002); Phys. Rev. A {\bf 67}, 033603 (2003).
\bibitem{Cla}
E. A. Donley,  N. R. Claussen, S. T. Thompson, and C. E. Wieman, Nature \textbf{417}, 529 (2002).
\bibitem{Du}
R. A. Duine, and H. T. C. Stoof,
J. Opt. B : Quantum Semiclass. Opt.~\textbf{5}, S212 (2003).
\bibitem{Mahan}
G.D. Mahan, {\it{Many particle physics}} (Plenum, New York, 1990).
\bibitem{Chi}
C. Chin, A. J. Kerman,V. Vuletic, and  S. Chu, Phys. Rev. Lett. \textbf{90}, 033201 (2003).
\end{thebibliography}
\end{document}